**High-temperature weak ferromagnetism on the verge of a metallic state: Impact of dilute Sr-doping on $BaIrO_3$**

G. Cao, X.N. Lin, S. Chikara, V. Durairaj, and E. Elhami
Department of Physics and Astronomy
University of Kentucky, Lexington, KY 40506

The 5d-electron based $BaIrO_3$ is a nonmetallic weak ferromagnet with a Curie temperature at Tc=175 K. Its largely extended orbitals generate strong electron-lattice coupling, and magnetism and electronic structure are thus critically linked to the lattice degree of freedom. Here we report results of our transport and magnetic study on slightly Sr doped $BaIrO_3$. It is found that dilute Sr-doping drastically suppresses Tc, and instantaneously leads to a nonmetal-metal transition at high temperatures. All results highlight the instability of the ground state and the subtle relation between magnetic ordering and electron mobility. It is clear that $BaIrO_3$ along with very few other systems represents a class of materials where the magnetic and transport properties can effectively be tuned by slight alterations in lattice parameters.

The layered iridates $BaIrO_3$ [1-5] and $Sr_{n+1}Ir_nO_{3n+1}$ with n=1 and 2 [6-11] possess two unique features not found in other materials, namely, high-temperature weak ferromagnetism and an insulating ground state with phase proximity of a metallic state. These features defy common wisdom that 4d and 5d transition metal oxides should be much more conducting than their 3d counterparts because of the more extended d-orbitals that significantly reduce the Coulomb interaction. In fact, this extended characteristic significantly enhances crystalline field splittings and the d-p hybridization between the transition metal and the oxygen octahedron surrounding it. This, in turn, leads to strong electron-lattice coupling, which very often alters and distorts the metal-oxygen bonding lengths and angles, lifting degeneracies of $t_{2g}$ orbitals and thus precipitating possible orbital ordering. It is this feature that defines a high sensitivity of the ground state to the atomic stacking sequence and structural distortions. This characteristic is illustrated in recent studies on 4d-electron based ruthenates where magnetism and electric transport change drastically and systematically with structural dimensionality. Although the layered iridates distinguish themselves from the ruthenates by showing conspicuously low magnetic moments and yet high Curie temperatures, magnetic and transport behavior of these 5d-electron systems is also critically linked to the lattice degree of freedom, requiring only dilute impurity doping to tip the balance across the metal-insulator borderline accompanying drastic changes in magnetic ordering.

The $Ir^{4+}$ ($5d^5$) ions in these layered iridates are presumed to be in a low spin state with S=1/2 since the large extension of the 5d orbitals leads to large crystal field splittings and a reduced Coulomb repulsion [1-11]. $BaIrO_3$ is a nonmetallic weak ferromagnet with an exotic ground state. Recent experimental studies reveal evidence for a simultaneous onset of weak ferromagnetism and charge density wave (CDW) formation at Tc=175 K [4], an unexpected occurrence that is also supported by results of tight-binding band structure calculations, which show partially nested pieces on the Fermi surface [5]. The major experimental observations include a simultaneous transition in magnetization and resistivity at Tc that is, however, insensitive to magnetic field; non-Ohmic behavior below Tc, and a gap (~0.15 eV) formation at Tc [4]. In addition, the saturation moment, $\mu_S$=0.03 $\mu_B$/Ir, which is not achieved until 20 T for T= 5 K, is only 3% of the moment expected for localized spin S=1/2. Such a small $\mu_S$ is probably intrinsic due to d-p hybridization and small exchange splitting, rather than spin canting from a localized antiferromagnetic configuration [4, 5]. The paramagnetic effective moment, $\mu_{eff}$=0.13 $\mu_B$/Ir, is also small compared to 1.41 $\mu_B$/Ir anticipated for an S=1/2 system, yielding a ratio of $\mu_{eff}/\mu_S$=4.3 much larger than unity expected for localized electrons. While the small $\mu_{eff}$=0.13 $\mu_B$/Ir is believed to be due to slight shifting of the up- and down-spin bands by about 37 meV, the weak ferromagnetism is attributed to a sharp peak of density of states near the Fermi surface anticipated in the tight-binding band structure calculations [5], consistent with the Stoner criterion. The band structure calculations also conclude that $BaIrO_3$ should be considered as a weakly localized metal above Tc and that the Fermi surfaces contain strongly warped cylinder-like pieces aligned along the c-axis [5]. Its sister compound $SrIrO_3$, on the other hand, is a paramagnetic metal, an interesting exception to all other iridates that are typified by high-temperature weak ferromagnetism and non-metallic behavior. Both $BaIrO_3$ and $SrIrO_3$ are non-perovskite with a nine-layer hexagonal structure for the former [1,2] and a six-layer hexagonal structure for the latter [12].

The crystal structure of $BaIrO_3$ features three face-sharing $IrO_6$ octahedra forming $Ir_3O_{12}$ clusters that are vertex-linked to construct one-dimensional (1D) chains along the *c*-axis [1-5]. $BaIrO_3$ is isostructural to metallic $BaRuO_3$ [13, 14], but the monoclinic distortion in $BaIrO_3$ generates twisting and buckling of the cluster trimers that give rise to two 1D zigzag chains along the *c*-axis and a layer of corner sharing $IrO_6$ octahedra in the ab-plane, bringing about both 1D and 2D structural characteristics [1-5]. The structural distortion effectively localizes d-electrons, resulting in the non-metallic behavior in a wide temperature range of $1.7<T<900$ K [4]. That the physical behavior of $BaIrO_3$ is vastly different from that of $SrIrO_3$ and $BaRuO_3$ clearly indicates a ground state that is strongly coupled with the lattice degree of freedom. It is therefore expected that the electron correlation strength can in effect be controlled by tuning lattice parameters through dilute impurity doping, an approach known as bandwidth (W) control that modifies W without altering charge carriers, the crystal structure and thus Coulomb interaction. Here we report results of our recent transport and magnetic study on slightly Sr doped $BaIrO_3$, namely, $Ba_{1-x}Sr_xIrO_3$. It is found that dilute Sr-doping drastically suppresses Tc, and instantaneously leads to a nonmetal-metal transition at high temperatures. The results underline the instability of the ground state and the subtle relation between magnetic ordering and electron mobility. $BaIrO_3$ along with very few other systems, such as $V_2O_3$, $NiS_{1-x}Se_x$ [15], represents a class of materials where magnetism and transport are primarily driven by the lattice degree of freedom alone.

Single crystals were grown using flux techniques [4]. The crystals studied were examined with x-ray diffraction and EDS. The results suggest high quality of the crystals with uniform Sr doping. The magnetic and transport properties were measured using a Quantum Design MPMS LX with an added function for transport measurements.

Shown in Fig. 1 is the lattice parameters, *a-, b-, c*-axis and volume of the unit cell, as a function of Sr concentration. Without any structural phase transition, the lattice parameters decrease quite linearly with increasing Sr concentration because the Sr ion with an ionic radius of $r_{Sr}$=1.18 Å is smaller than the Ba ion with $r_{Ba}$=135 Å, basically consistent with the Vegard's law. These decreases in the lattice parameters expectedly relax the buckling of $IrO_6$, and, as a result, effectively increase the kinetic energy of 5d-electrons or the bandwidth. This is the key feature of this system.

Shown in Fig. 2 is the magnetization, M, for the c-axis as a function of temperature (a) and magnetic field at T=25 K (b) for a few representative Sr concentrations (The c-axis is the magnetic easy axis. Magnetic moment for the ab-plane is about 15% smaller [4]). As can be seen in Fig.2a, the weak ferromagnetic transition decreases rapidly with increasing Sr concentration and vanishes at x>0.23. The magnetic ordering for x=0.23 is suppressed to about 17 K but is rounded and less defined (see the inset). Fitting the data for x=0.23 into the Curie-Weiss law yields a negative Curie-Weiss, $\Theta_{CW}$ =-17 K, suggesting a crossover to an antiferromagnetic exchange coupling. The saturation moment $\mu_S$ also decreases rapidly with the Sr concentration, and eventually becomes zero for x>0.23 as shown in Fig.2b. The magnetization M for x<0.23 is very hysteretic, i.e., when the field is reduced to B=0 T, M remains essentially unchanged from the saturation value. In addition, that the magnetic susceptibility above Tc for all x is nearly temperature independent suggests Pauli paramagnetic behavior, a characteristic for a metal.

The changes in the magnetic behavior are accompanied by rapid changes in transport properties. Illustrated in the Fig.3 is the electrical resistivity $\rho_{ab}$ for the basal plane as a function of temperature for 2≤T≤350 K (Behavior of resistivity for the c-axis as a function of x is similar). Fig. 3a shows $\rho_{ab}$ for the pure host (x=0) at B=0 and 7 T. There are two remarkable features, namely, the abrupt slope change of $\rho_{ab}$ at Tc=175 accompanying the occurrence of weak ferromagnetism, and essentially unchanged $\rho_{ab}$ and Tc at applied magnetic field B=7 T (These features occur in all samples measured although temperature dependence of ρ below Tc may slightly vary from sample to sample). The application of a magnetic field to a ferromagnet is normally expected to cause broadening of Curie temperature and magnetoresistance. That $\rho_{ab}$ changes sharply at Tc and yet shows no sensitivity to the magnetic field as evidenced in Fig. 3a suggests that the magnetism and transport are coupled through the lattice degree of freedom. While the slope change of ρ at Tc may be attributed to the formation of a CDW [4,5], it cannot be ruled out a possible presence of orbital ordering below Tc. In perfectly symmetric octahedra, the states are split into a lower $t_{2g}$ triplet and an excited $e_g$ doublet. Only the $t_{2g}$ orbitals play a role here, because the exchange between the 5d electrons is too small to align the spins into an S=5/2 (first Hund's rule). The orbital ordering may be facilitated by lifting the degeneracy of the $t_{2g}$ orbitals via a lattice distortion of the Jahn-Teller type. Although subject to more investigations, there is some evidence indicating such a structural change at Tc [4].

Indeed, the dilute Sr substitution for Ba instantly results in metallic temperature dependence of $\rho_{ab}$ at high temperatures with a drastically reduced magnitude of $\rho_{ab}$. As can be seen in Fig. 3b and c, the metallic temperature dependence of $\rho_{ab}$ clearly becomes stronger with increasing x. The unusual temperature dependence of $\rho_{ab}$ for x=0.06 shown in Fig.3 b is particularly interesting, suggesting an electronic state verging on the metallic borderline; $\rho_{ab}$ immediately shows metallic behavior near Tc=145 K, in contrast to that of the pure compound where the onset of Tc is accompanied by the more insulating behavior as shown in Fig.3a. The resistivity minima at low temperatures shown in Fig.3b and c, for example, 30 K for x=0.06, is somewhat similar to that of the pure compound and is not evidenced in M. The ubiquitous nature and its apparent insensitivity to the application of a magnetic field, in contrast to the behavior expected from a Kondo system, suggests that it is a fundamental property of the lattice rather a magnetic or spin fluctuation effect. Similar behavior has been observed in some magnetic metals with disordered lattice such as $Fe_{40}Ni_{20}B20$ [16]. For x=0.23, the system becomes fully metallic with $\rho_{ab} \propto AT^2$ below 50 K expected for a Fermi liquid. The coefficient, $A \propto \mu_S^{-1}$ expected for a weak ferromagnet, is estimated to be $1.5x10^{-7}$ Ω cm, comparable to those for weak ferromagnets such as $ZrZn_2$ and $Sc_3In$ ($5x10^{-7}$ Ω cm) [17, 18] but much larger than those for ferromagnets such as $SrRuO_3$ ($9.3x10^{-9}$ Ω cm) [19] and Fe ($1x10^{-11}$ Ω cm) [17]. Markedly, $\rho_{ab}$ for x=0.23 is extraordinarily linear in temperature for 50 <T<350 K, which could be attribute to the electron-phonon interaction, and it becomes eventually saturated at T>450 K (not shown). In fact, $\rho_{ab}$ for all smaller x tends to saturate, as evidenced in Fig. 3c. The saturation of $\rho_{ab}$ at high temperatures is more likely a signature of a breakdown of the Boltzmann theory although it is also theoretically anticipated for weak ferromagnets due to the increase in spin fluctuations [15]. The absence of magnetic field dependence of ρ in $Ba_{1-x}Sr_xIrO_3$ suggests an insignificant role of spin fluctuations in scattering process. Generally, in nonlocal spin systems such as $ZrZn_2$, spin fluctuations are small and extended in real space and localized in *q*-space.

It is argued based on the band structure calculations that $BaIrO_3$ is actually a weakly localized metal above 175 K in spite of the nonmetallic temperature dependence of resitivity and that the $t_{2g}$-block band density of state has a sharp peak at the Fermi level [5]. In a normal metal, the charge carriers are delocalized and their wave functions extend over the whole crystal. Since the carriers experience inelastic scattering against phonons, defects and other electrons, both the resistivity of a normal metal and the mean free path become finite. A system such as $BaIrO_3$ with partially filled bands is subject to electron localization induced by random potential associated with lattice defects and electron-phonon coupling as found in polaronic states [5]. The weak localization above Tc=175 K in $BaIrO_3$ could be attributed to the disorder that are low-energy vibrations associated with the twisting and bending of the $Ir_3O_{12}$ trimer units around the Ir-O-Ir bridges [5]. The nonmetallic behavior at low temperatures is, at least, due partially to the formation of the gap of 0.15 eV found below Tc [4]. The fact that rather small lattice changes can cause such drastic alterations of electronic properties demonstrates how weak the localization is and how subtle the electronic and magnetic ground states are in $BaIrO_3$. Apparently, these ground states are delicately coupled to the lattice. This point is further manifested in Fig. 4 where M (T) and ρ (T) (left scale) for the ab-plane for 7 % Ca doped $BaIrO_3$ is presented. The Ca ($r_{ca}$=1.00 Å) ion too small for substituting for Ba ($r_{Ba}$=1.35 Å) causes severe structural distortions, resulting in a much more insulating state with an increase of more than 7 orders of magnitude in ρ and vanishing weak ferromagnetism (see Fig.4).

It is remarkable that the ratio of $\mu_{eff}/\mu_S$ vs Tc, shown in Fig.5a, is qualitatively consistent the Rhodes-Wohlfarth plot [15]. $\mu_{eff}/\mu_S$ ranging from 3.5 to 20 with decreasing Tc illustrated in Fig. 5a suggests an intermediate regime (which may be described by Stoner model) between a ferromagnetic insulating state ($\mu_{eff}/\mu_S$=1) and a paramagnetic metallic state ($\mu_{eff}/\mu_S=\infty$) where a large number of electrons moves out of the shell and become itinerant. Indeed, the transport and magnetic behavior clearly exhibits parallel changes with the variation of Sr concentration (see Fig.5b), that is, electrons become more itinerant as spin polarization decreases, indicating that the mobility of electrons is susceptible to magnetic ordering. There is noticeably a narrow range where both the metallic behavior and weak ferromagnetism coexist, highlighting the subtle balance between localized moments and itinerant electrons. While the transport behavior of the system obviously traces the weak ferromagnetic ordering, it is clear that the magnetism and the electronic structure are so critically coupled to the lattice degree of freedom that small perturbations in lattice alone can precipitate drastic changes in bandwidth, and thus the ground state.

$BaIrO_3$ shares a number of unique features with other Iridates such as the layered $Sr_2IrO_4$ and $Sr_3Ir_2O_7$. They are all characteristically insulating with unusual temperature dependence of resistivity and negative differential resistivity or non-ohmic behavior [4, 10, 11]. These iridates are all weakly ferromagnetic with high magnetic ordering temperatures (Tc=240 K and 285 K for $Sr_2IrO_4$ and $Sr_3Ir_2O_7$, respectively) and yet only a fraction of the expected ordered moment ($\mu_S$=0.14 $\mu_B$/Ir and 0.04 $\mu_B$/Ir for $Sr_2IrO_4$ and $Sr_3Ir_2O_7$, respectively) [10,11]. In addition, all these materials often show a simultaneous transition in magnetization and resistivity, but none displays sensitivity of Tc to high magnetic fields and detectable magnetoresistance near and below Tc except at very low temperatures [10, 11]. It is striking that all their counterparts in the ruthenates ($BaRuO_3$, $Sr_2RuO_4$, and $Sr_3Ru_2O_7$) are metallic or even superconducting and that all these ruthenates are less structurally distorted. Clearly, the lattice degree of freedom play a critical role in determining the ground state in these systems with extended electron orbitals. Without doubt, this class of the materials, which are often characterized by overwhelming inter-atomic interactions, presents new challenges and merits systematic investigations.

Acknowledgement: This work was supported by an NSF grant DMR-0240813.

Captions:

Fig.1. The lattice parameters, *a-, b*-axis (a) and *c*-axis and volume of the unit cell (b), as a function of Sr concentration.

Fig.2. Magnetization, M, for the c-axis as a function of temperature at B=0.01 T (a) and magnetic field at T=25 K (b) for a few representative Sr concentrations. Inset: M vs T for x=0.23.

Fig.3. Electrical resistivity $\rho_{ab}$ for the basal plane as a function of temperature for $2 \leq T \leq 350$ K for x=0 (a), x=0.06 (b) and x=0.12, 0.14 and 0.23 (c).

Fig.4. M (T) and ρ (T) (left scale) for the ab-plane for $Ba_{0.93}Ca_{0.07}IrO_3$.

Fig.5. (a) The ratio of $\mu_{eff}/\mu_S$ as a function of Curie temperature Tc; Tc and approximate metal-insulator transition as a function of Sr concentration x. Note that PM stands for paramagnetism and WFM for weak ferromagnetism.

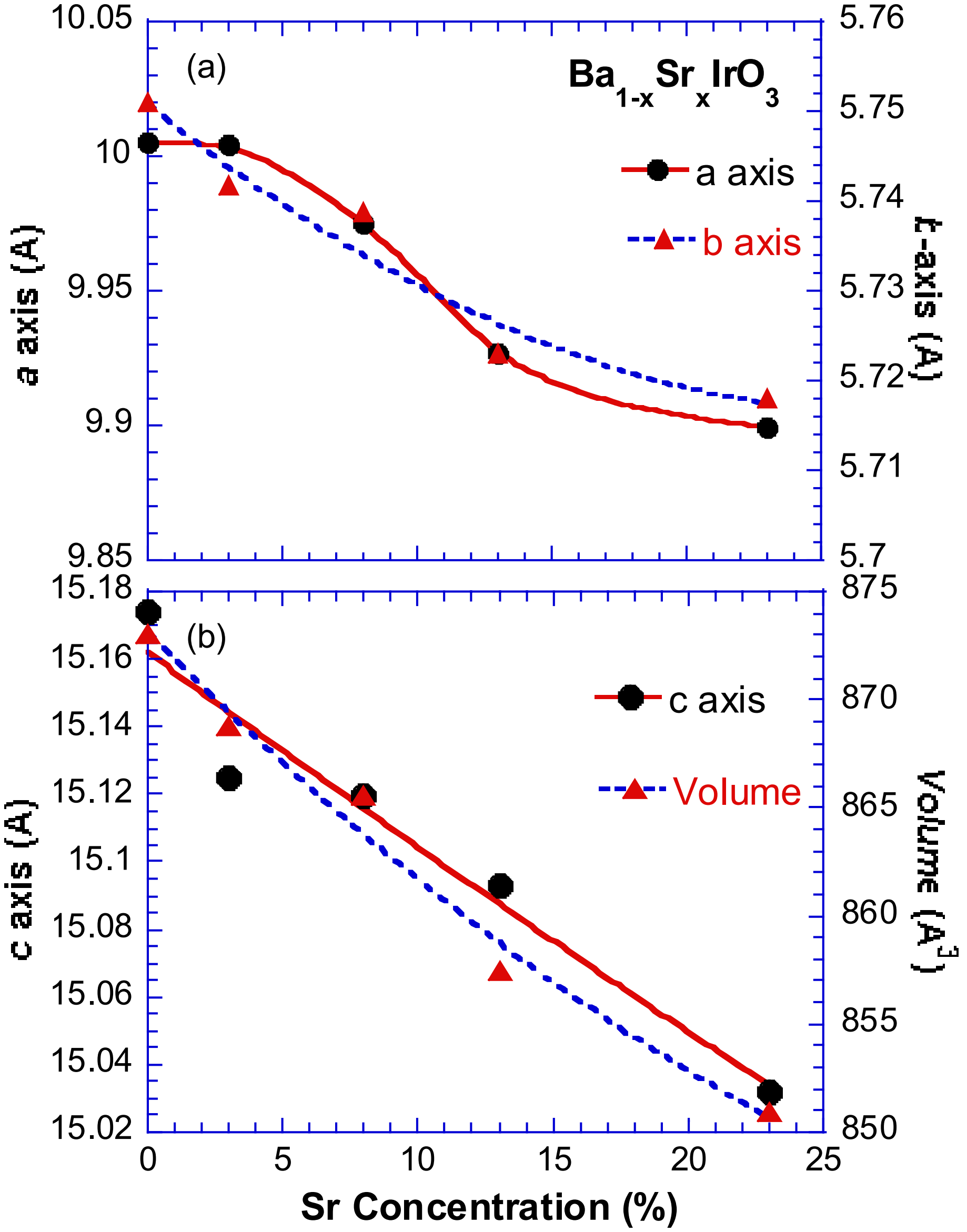

Fig.1, Cao

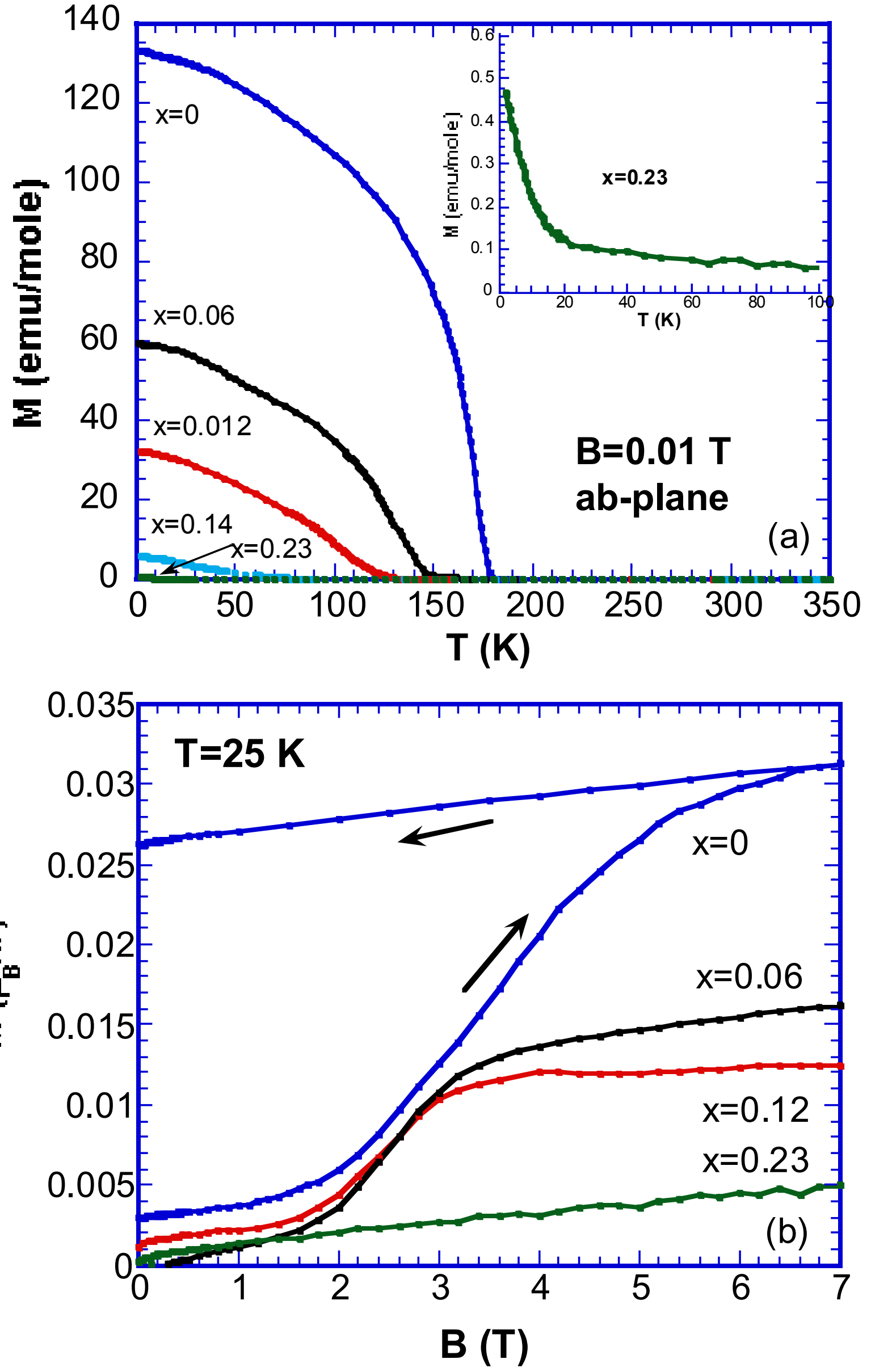

Fig.2, Cao

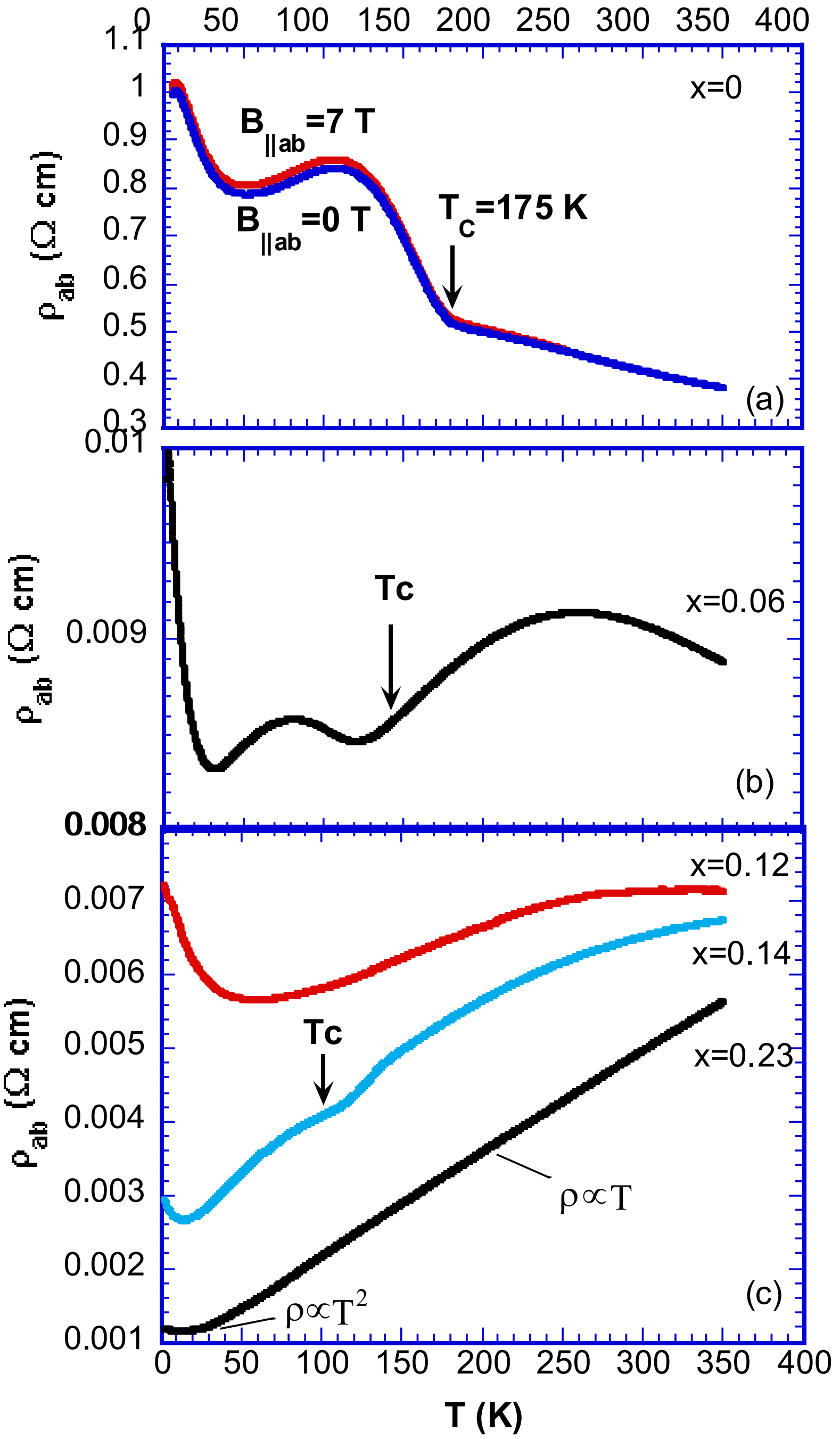

0 50 100 150 200 250 300 350 400
1.1
1
0.9
0.8
0.7
0.6
0.5
0.4
0.3
x=0
B||ab=7 T
B||ab=0 T
Tc=175 K
(a)
ρab (Ω cm)
0.01
0.009
0.008
Tc
x=0.06
(b)
ρab (Ω cm)
0.007
0.006
0.005
0.004
0.003
0.002
0.001
x=0.12
x=0.14
x=0.23
Tc
ρ∝T
ρ∝T2
(c)
ρab (Ω cm)
0 50 100 150 200 250 300 350 400
T (K)

Fig.3,Cao

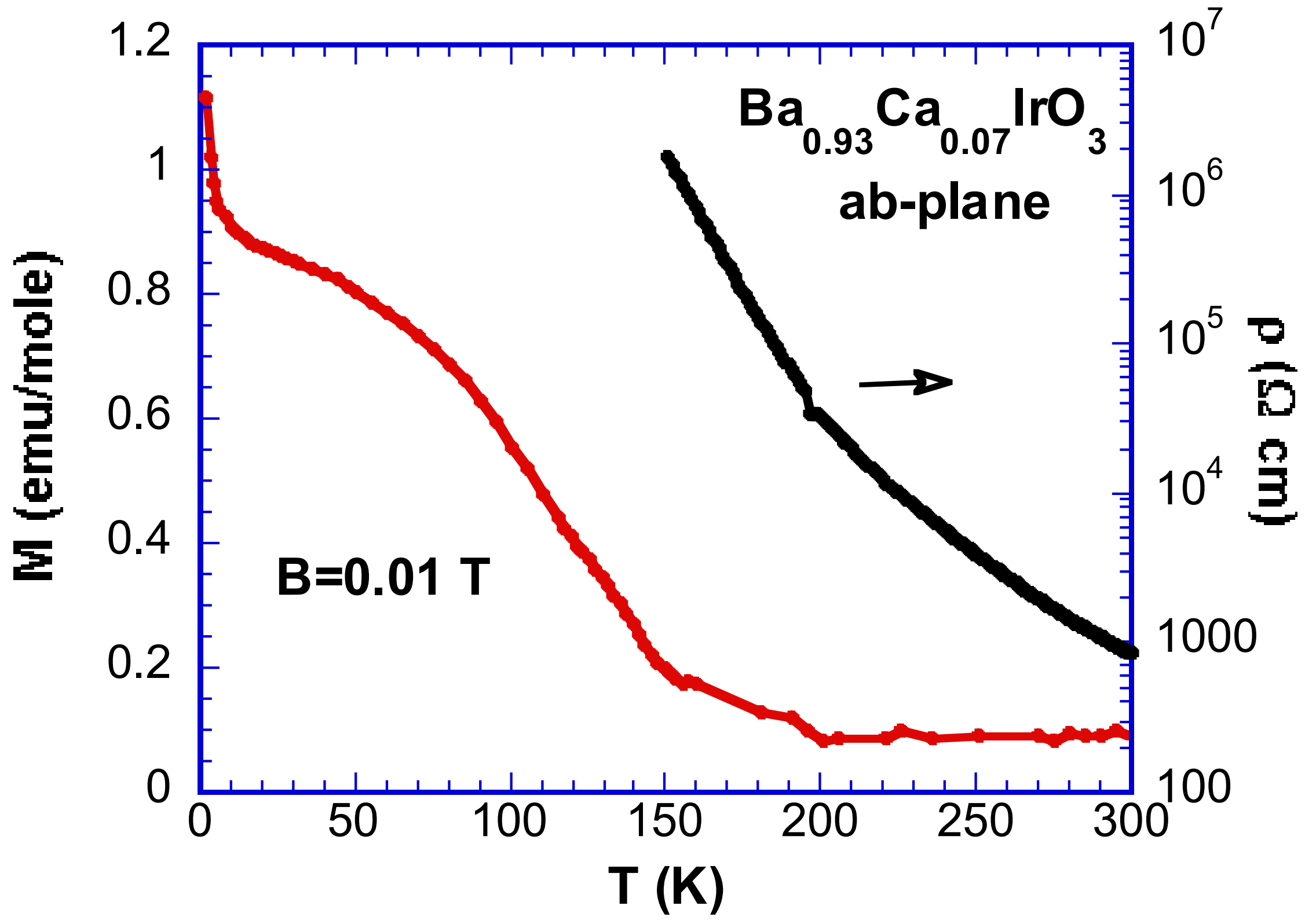

Fig.4, Cao

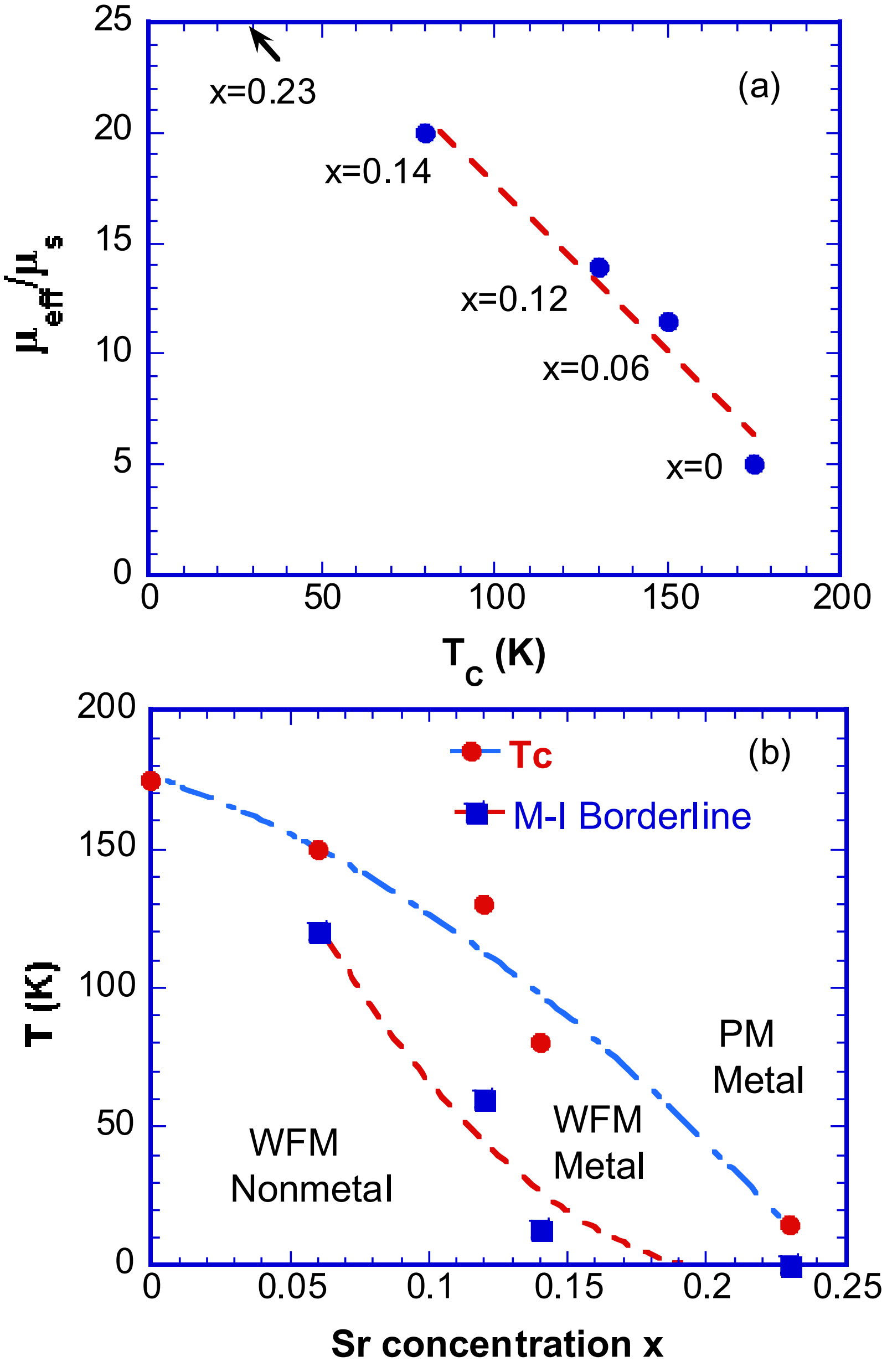

Fig.5, Cao